\documentclass[%
a4paper,
prd,
twocolumn,
superscriptaddress,
preprintnumbers,
nofootinbib,
nobibnotes,
amsmath,amssymb,
aps,
floatfix
]{revtex4-2}
\usepackage{amsfonts,graphics,color}
\usepackage[english]{babel}
\usepackage{graphicx}
\usepackage{amsmath}
\usepackage{dcolumn}
\usepackage{bm}
\usepackage{multirow}
\usepackage{xcolor}
\usepackage{comment}
\usepackage{ marvosym }
\definecolor{xlinkcolor}{cmyk}{1,1,0,0}
\usepackage[bookmarks=true, pdfnewwindow=true, colorlinks=true, linkcolor=xlinkcolor, citecolor=xlinkcolor, filecolor=xlinkcolor, urlcolor=xlinkcolor, final=true]{hyperref}

\usepackage[normalem]{ulem}

\begin{document}

\preprint{INR-TH-2026-003}

\title{Constraints on millicharged particles from thunderstorms on the Solar system planets
}

\author{Ekaterina Dmitrieva}
\email[\textbf{e-mail}: ]{edmitrieva@inr.ru}
\thanks{corresponding author}
\affiliation{Institute for Nuclear
Research of the Russian Academy of Sciences, 60th October Anniversary Prospect 7a, Moscow 117312, Russia}
\affiliation{Faculty of Physics, Moscow State University, Leninskiye Gory, 119991 Moscow, Russia}

\author{Petr Satunin}
\email[\textbf{e-mail}: ]{satunin@ms2.inr.ac.ru}
\affiliation{Institute for Nuclear
Research of the Russian Academy of Sciences, 60th October Anniversary Prospect 7a, Moscow 117312, Russia}
\date{April 2026}

\begin{abstract}
    
 We investigate the production of millicharged particles (mCPs) by the Schwinger mechanism in thunderstorms in the atmospheres of different planets in the Solar system. We consider a thundercloud as a giant capacitor that can be discharged in two ways: either by lightnings or by mCP production. Taking into account the observation of lightning strikes, we establish the constraints on the charge and mass of mCPs. We examine two types of cloud configurations: a simple arrangement of two clouds, and a more complex layered structure that gives rise to potential wells. In the latter case, we take into account the effects of Bose enhancement for scalar mCPs, and Pauli blocking for fermionic ones. 
 We use the observational data of planetary atmospheres  obtained by satellite missions to establish constraints on the charge and mass of mCP particles. The best constraints came from the observation of thunderstorms in Saturn's atmosphere under an assumption of layered cloud structure:
$q > 10^{-24}$ for bosonic mCPs. These constraints for bosons are, to the best of our knowledge, the best in the literature.

\end{abstract}

\maketitle

\section{Introduction}

In the last half century, there has been great progress in the observation of the atmospheres of Solar System planets by satellite missions. The first satellite missions that traveled near Jupiter were Pioneers 10 and 11 \cite{Pioneer10} which made the first observations of Jovian atmosphere \cite{wallace1974thermal}; and Voyagers 1 and 2 \cite{kohlhase1977voyager, cook1979first, becker2020small, scarf1981upper}. The first spacecraft that reached the Jovian orbit was Galileo \cite{gierasch2000observation, seiff1996structure} (1995 - 2003). It released a probe that made the first direct measurements of the Jovian atmosphere. 

The current spacecraft mission is JUNO \cite{JUNOwebsite, bolton2017juno, wong2026radio}, which arrived to a stationary polar orbit near Jupiter in 2016. It has performed several important discoveries regarding the structure of the Jovian atmosphere, including the observation of lightnings. Notably, the space and energy scale of atmospheric processes in the Jovian atmosphere are, in some cases, an order of magnitude greater than those of the processes on Earth \cite{wong2026radio}.  

The lightning on Saturn was discovered by Voyager 1 \cite{doi:10.1126/science.212.4491.239}, which detected high-frequency radio emissions in 1980. This was the first evidence of lightning on Saturn, and these signals were named Saturn Electrostatic Discharges (SEDs). The SED was investigated in detail by Cassini and Voyager 2 \cite{fischer2008atmospheric}. Them are characterized by energies $\sim10^{12}\mbox{J}$, which is one of the most powerful lightning in the Solar System \cite{luque2014coupling}. The first optical detection of lightning on Saturn was in 2010 \cite{dyudina2010detection}. 

Our most accurate data on the planetary atmospheres come from Venus \cite{seiff1985models}. However, relatively little is known about its lightning activity. The Venera 11, 12 and Pioneer Venus detected very low frequency (VLF) signals corresponding to lightning in Venusian atmosphere in 1978 \cite{ksanfomaliti1980discovery, ksanfomaliti1979electrical, scarf1980lightning, williams1983planetary}. Optical signals from lightning were detected by Venera 9 in 1975 \cite{krasnopolsky1980lightnings} and by ground-based telescopes in 1995 \cite{HANSELL1995345}. 


Besides the investigation of the structure of these planets, the observations of processes on a larger scale than those on Earth, as well as observations of giant atmospheres, can be a very good probe for some hypotheses in fundamental physics (so-called new physics). In this letter, we concentrate  on the popular extension of the Standard model of particle physics, called  millicharged particles (mCP). 

These are particles with a small mass and a tiny electric charge, much smaller than the electron charge \cite{IGNATIEV1979315,Okun:1982xi, HOLDOM1986196}. 
Millicharged particles appear in extensions of the Standard Model \cite{HOLDOM1986196, Batell:2005wa, Abel:2003ue, Abel:2008ai}, grand unification theories \cite{Pati:1973uk, Georgi:1974my, Preskill:1984gd}, and string theory \cite{Wen:1985qj, Shiu:2013wxa}. 
They  are also considered as as candidate to be dark matter (DM) particles \cite{Brahm:1989jh, Feng:2009mn, Cline:2012is, Gong:2025xsd} or be as one of particles of a wider dark sector \cite{Okun:1982xi, HOLDOM1986196, Jaeckel:2021xyo, Shiu:2013wxa, Brahm:1989jh, Feng:2009mn, Cline:2012is}. mCP dark matter can be bosonic or fermionic; there are strong constraints on light fermionic mCP DM due to Pauli principle \cite{Rubakov:2017xzr}. Besides, there are searches for mCPs production in \cite{Gninenko:2025aek, deMontigny:2023qft, Fiorillo:2024upk, Davidson:2000hf, Berlin:2021kcm, Chang:2018rso, Lepidi:2007vnd, Cruz:2022otv, Arza:2025cou, Dmitrieva:2025ohn, Berlin:2024dwg} without an assumption for mCP to be a DM particles.

Millicharged particles can be produced by electromagnetic interactions in perturbative or non-perturbative processes.  An interesting non-perturbative channel is Schwinger pair-production \cite{schwinger1951gauge} — the spontaneous production of charged particle-antiparticle pairs in a strong electric field. The Schwinger process for electrons is exponentially suppressed until the electric field strength reaches the critical value  $E_{cr}=m^2/e \approx 1.3\times 10^{18}~\mbox{V/m}$. This is an extremely large value that can hardly be realized  in Nature. However, for mCPs with a very small mass instead of electrons, the critical field may be of reasonable value even for a quite small charge.

In the context of mCPs, Schwinger pair production was considered in \cite{doi:10.1142/S0217732314500540,Berlin:2020pey, Berlin:2024dwg,Kouvaris:2025tom}: a large electric field can be generated in laboratory capacitors\cite{doi:10.1142/S0217732314500540}, in superconducting radiofrequency cavities \cite{Berlin:2020pey}, in the polar gaps of pulsars \cite{Kouvaris:2025tom}, and in the Earth's atmospheric processes, including thunderstorms \cite{Berlin:2024dwg}.

The authors of  \cite{Berlin:2024dwg}  considered the case of Bose enhancement of the Schwinger process in a potential well and established the constraints on Bose mCPs parameters using Earth's atmosphere data including thunderstorms. Thunderstorms are capable of producing intense electric fields. A thundercloud behaves like a capacitor on the scale of kilometers: electric charges gather within it, increasing the field strength until a lightning strike releases the stored charge. If mCPs are present in Nature, discharge could occur by an alternative channel of mCP production, establishing constraints on mCP parameters. The authors of \cite{Berlin:2024dwg}  considered the case of bosonic mCPs in potential well between layered thunderclouds at Earth. 
In this letter we extend this idea in several ways: consider fermionic mCP in addition to bosonic ones, two cloud structure in addition to the layered one, and data from thunderstorms on other planets in the Solar system \cite{aplin2017lightning, yair2008updated, nagy2008comparative, aplin2017electrical} that are more intense and extended than the terrestrial ones. So, we expect better constraints on bosonic mCPs.  

\section{Schwinger production of millicharged particles in a thundersorm}\label{sec_schwinger}

\begin{figure}
    \centering
    \includegraphics[width=0.5\linewidth]{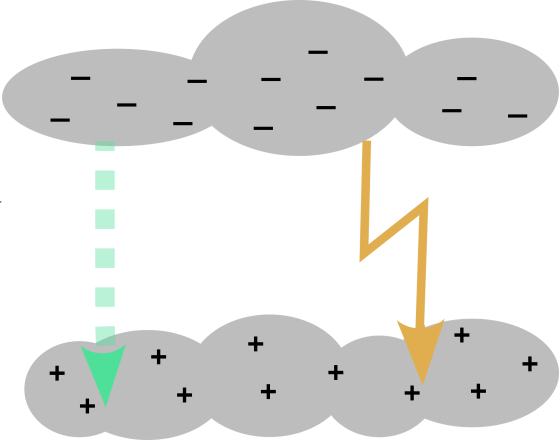}
    \caption{Two cloud thunderstorm. Discharge of thunderclouds due to mCP production (left) or lightning (right).}
    \label{fig:Clouds1}
\end{figure}

In this Section, we consider Schwinger pair production of mCPs between two charged thunderclouds of large area and distance $L$ between the clouds, which can be treated as a giant capacitor with a quasiuniform constant electric field inside. The scheme of two clouds is illustrated in Fig.~\ref{fig:Clouds1}. Two oppositely charged clouds  discharge due to lightning if the electric field between the clouds exceeds a threshold value. If mCPs exist in Nature, they provide an alternative channel for cloud discharge due to Schwinger pair production.    

The width  (probability per unit volume per unit time) of Schwinger pair production of charged particles with a charge $q$ and  mass $m$ in an electric field $E$ is given, for scalar and fermion particles, consequently\footnote{Natural system of units $\hbar = c= 1$ is used.} \cite{schwinger1951gauge, NIKISHOV1970346},
\begin{align}
   & \Gamma_S = \frac{(qE)^2}{(2\pi)^3} \sum_{n=1}^{\infty} \frac{(-1)^{n+1}}{n^2} e^{-\frac{\pi m^2}{qE}n}, \label{eq:Schwinger-scalar}\\
    & \Gamma_F = \frac{(qE)^2}{4\pi^3} \sum_{n=1}^{\infty} \frac{1}{n^2} e^{-\frac{\pi m^2}{qE}n}. \label{eq:Schwinger-fermion}
\end{align}

Non-perturbative over arbitrary $q$ calculations \cite{schwinger1951gauge, NIKISHOV1970346} show that the Schwinger series for bosons \eqref{eq:Schwinger-scalar} are sign-alternating while the sign for fermions \eqref{eq:Schwinger-fermion} does not alternate with $n$. Besides, there is an additional factor $2$ in the pre-exponential factor of eq.~\eqref{eq:Schwinger-fermion} due to two spin states for fermions.

In the regime $qE \ll m^2$, even the first term in the series \eqref{eq:Schwinger-scalar},\eqref{eq:Schwinger-fermion} is exponentially suppressed, so the Schwinger process effectively does not hold. In the opposite limit $qE \gg m^2$ called supercritical, the series \eqref{eq:Schwinger-scalar},\eqref{eq:Schwinger-fermion} can be taken explicitly, providing the result
\begin{equation}
\label{eq:Schwingwer-supercritical}
     \Gamma_S = \frac{(qE)^2}{96\pi}, \qquad \Gamma_F = \frac{(qE)^2}{24\pi}. \qquad (qE \gg m^2)
\end{equation}

\begin{figure}
    \centering
    \includegraphics[width=0.85\linewidth]{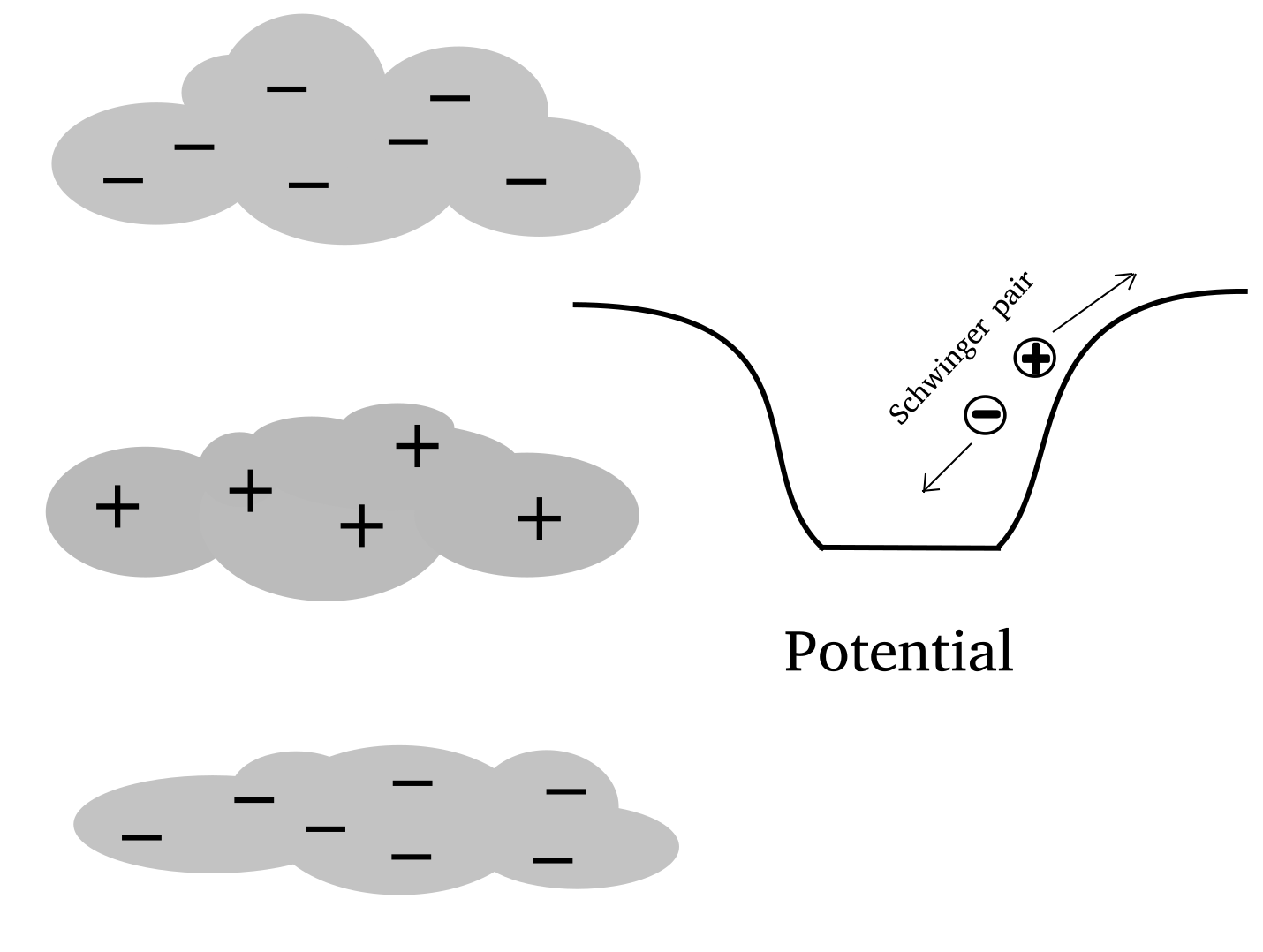}
    \caption{The layered structure of charged thunderclouds (left) and potential well for produced mCPs (right). Negatively charged mCP  from the Schwinger pair remain in the potential well.}
    \label{fig:Clouds2}
\end{figure}
The produced charged mCPs create an electric current $J_{\rm mc}$ related to the discharge of the cloud capacitor (see Fig.~\ref{fig:Clouds1}).  The current reads,

\begin{equation}
\label{eq:mc-current}
   J_{mc} \simeq 2\times q\times \Gamma_{S(F)} \times L^3, 
\end{equation}
where $\Gamma_{S(F)}$ is related to the scalar/fermion Schwinger width \eqref{eq:Schwingwer-supercritical}. The capacitor volume is estimated as $L^3$, and factor $2$ is related to the two independent currents of positive and negative mCPs in opposite directions. 

If the lightning is observed, the capacitor discharge is due to fast lightning, not due to continuous discharge from mCP production (see Fig.~\ref{fig:Clouds1}). Thus, the following relation between currents reads:
\begin{equation}
\label{eq:currents}
    J_{\rm mc} <  J_{\rm lightning} \times \frac{t_{{\rm lightning}}}{t_{\rm pause}},
\end{equation}
where $t_{\rm lightning}$ is a short duration of one lightning strike, $J_{\rm lightning}$ is the amplitude of the electric current, and $t_{\rm pause}$ is the average period between lightning strikes. 
Combining eqs.~\eqref{eq:Schwingwer-supercritical}, \eqref{eq:mc-current}, and \eqref{eq:currents}, one obtains the formula for the constraint on the mCP charge from the observation of a thunderstorm, 
\begin{equation}
\label{master-no-bose}
    q > \left( \frac{48\pi}{s\times E^2\times L^3} \times  J^{\rm amp}_{\rm lightning} \times \frac{t_{\rm lightning}}{t_{\rm pause}}  \right)^{1/3}.
\end{equation}
Here we assumed $q \gg m^2/E$, and introduced the spin factor $s=1$ for scalars and $s=4$ for fermions. 

\section{Bose enhancement effect for scalar millicharged particles}\label{sec_enhance}

For mCPs with Bose statistics, an additional enhancement factor may appear if the cloud structure consists of layers of clouds with opposite electric charges, which form several electrostatic potential wells (see Fig.~\ref{fig:Clouds2}). 
Bosonic mCPs of a certain charge are captured in these wells, forming a Bose-Einstein condensate while the mCPs of the opposite charge  escape from the potential well (see Fig.~\ref{fig:Clouds2}). The stimulated production of mCPs in the presence of the condensate grows exponentially with time \cite{klein1978bose}.

The concentration of mCPs with a fixed wavenumber, or momentum $p$ in the potential well — the mCP phase density  — is governed by the Boltzmann equation,
\begin{equation}
\label{eq:fp}
    \dot{f}_p = \Gamma'_p \left(  1 + f_p\right),
\end{equation}
which has an exponentially growing solution,
\begin{equation}
    f_p = e^{\Gamma'_p t} - 1.
\end{equation}
Here $\Gamma'_p$ is the rate for the collection of mCP produced by the Schwinger process in the potential well. 

The momentum spectrum for scalar Schwinger production reads \cite{NIKISHOV1970346, Cohen:2008wz}: 
\begin{equation}
\label{Nikishov}
    P = \frac{V}{(2\pi)^3} \int dp_3 \int d^2p_\perp \log \left( 1 + e^{-\frac{\pi(m^2+ p_\perp^2)}{qE}}\right).
\end{equation}
Here, the integration reads:
\begin{equation}
    \int dp_3 \to qET, \qquad \int d^2p_\perp   e^{-\frac{\pi p_\perp^2}{qE}n} = {\frac{qE}{\pi n}}.
\end{equation}
The mCP collected to the potential well from the distance $L$ so we can consider $T \sim L$. Finally, the rate $\Gamma'_p$ does not depend on $p$ if $p < \sqrt{qE}$,
\begin{equation}
    \Gamma'_p = L^{-1}.
\end{equation}

Since the considered volume is a cube $L^3$, all momenta should be larger than $L^{-1}$. The length scale corresponding to this process is  the Compton wavelength $d_C\sim 1/m_\chi$. Pair production is exponentially suppressed for $d\gtrsim d_C$ and is not suppressed if $d\lesssim d_C$. As we see in \eqref{Nikishov}, the minimal momenta play the role of mass in the exponent if $m < L^{-1}$. Thus, the Schwinger process is not suppressed if
\begin{equation}
\label{necessary}
    q > \frac{\pi\, \mbox{max}(m^2, L^{-2})}{E}.
\end{equation}
For the case of large mass  $m \gg L^{-1}$ the constraint  does not depend on $L$, becoming $q\gtrsim m^2/E$.

The overall charge density of the mCP condensate accumulated in the potential well is as follows:
\begin{equation}
\label{rho-mc}
    \rho_{\rm mc} = q \times \int d^3p \; f_p = q \,(qE)^{2}L\left(  e^{L^{-1}\; t} - 1 \right).
\end{equation}
The exponential growth stops when the overall Coulomb charge of the mCP condensate becomes an essential part ($\sim 0.5$) of the charge of the thundercloud. A rough estimation for the mCP condensate charge reads $Q_{\rm mc} = \rho_{\rm mc}\times L^3$, while  the charge in the thundercloud is $J^{\rm amp}_{\rm lightning} \times t_{\rm lightning}$. The discharge due to mCP condensate production should be faster than the pause between lightnings $t_{pause}$. 
Taking $t_{\rm pause} \sim 1$ sec and $L \sim 1$ km, in \eqref{rho-mc} one obtains an enormous exponential factor $e^{3\times 10^5}$. Thus, if the necessary condition  for the Schwinger effect \eqref{necessary} is satisfied and the potential well in the cloud structure really exists, the bosonic mCP is produced in a resonantly exponentially growing way.  

Thus, the necessary condition \eqref{necessary} in the case of Bose enhancement is also sufficient. So, we establish the constraints on mCP parameters from the observation of thunderstorms on different planets from this equation.

\section{The Pauli blocking effect for fermionic millicharged particles}\label{sec_pauli}

For mCP with Fermi statistics in a potential well (see Fig.~\ref{fig:Clouds2}) there is a maximal number of states \cite{landau2013statistical}. Considering momentum range $L^{-1}\lesssim p\lesssim \sqrt{qE}$ and taking into account degeneracy for two spin states, one obtains the number of quantum states,  
\begin{equation}
\label{eq:Fermi_number}
    N_F=\frac{V}{\pi^2}\int^{\sqrt{qE}}_{L^{-1}}p^2dp=\frac{L^3(qE)^{3/2}-1}{3\pi^2},
\end{equation}
where the volume is $V=L^3$.

On the other hand, we can estimate the number of mCPs, that are necessary to discharge the cloud capacitor:
\begin{equation}
\label{eq:discharge number}
    N_Q=\frac{Q}{q}=\frac{E\varepsilon S}{q},
\end{equation}
where $\varepsilon$ is dielectric permittivity, and 
$S=L^2$ is area of the capacitor plates. The constraint on $q$ can be obtained solving equation $N_F = N_Q$ with respect to q.

\section{Constraints on mCP parameters from observational data}\label{sec_estimation}

\begin{figure}[h]
    \centering
    \includegraphics[width=1.01\linewidth]{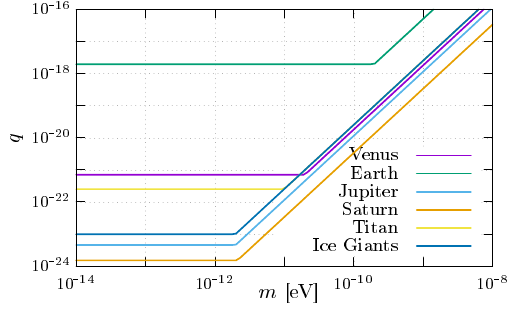}
    \caption{The constraints on the charge $q$ of bosonic mCP from the thunderstorms in the atmospheres of Solar system planets, in case of Bose enhancement. The sloping line correspond the bound value $qE=m^2$ for Schwinger suppression.}
    \label{fig_TS}
\end{figure}

\begin{figure}[h]
    \centering
    \includegraphics[width=1.01\linewidth]{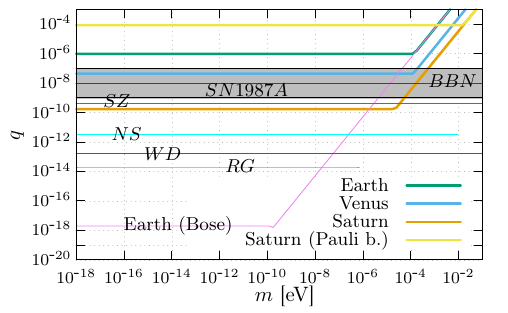}
    \caption{The constraints on the charge $q$  from the thunderstorms in the atmospheres of Solar system planets, in case of non-layered cloud structure (the corresponding lines refers to Earth, Venus and Saturn), in case of fermionic mCP and layered cloud structure (Pauli blocking effect), 
    and current experimental constraints: SN1987A \cite{Chang:2018rso, Davidson:2000hf, Davidson:1993sj}, Big-bang nucleosynthesis (BBN) \cite{Davidson:1993sj, Davidson:2000hf}, Earth(Bose) is constraints from Earth thunderstorms with Bose-enhancement (see Fig.\ref{fig_TS}), RG is red giant and WD is white dwarf \cite{Davidson:2000hf}, SZ is CMB observations of the Sunyaev-Zel’dovich effect \cite{Burrage:2009yz}, NS is neutron stars \cite{Korwar:2017dio}}.
    \label{fig_TS_noBose}
\end{figure}

Here we constrain the mCP charge $q_{Pl}$\footnote{Further in this section, we replace the index ``Pl'' with the first letter of the corresponding planet name: ``E'' for Earth, ``V'' for Venus, etc.} and mass $m_{Pl}$ from thunderstorms in the atmosphere on different planets in the Solar system. 


\textbf{Earth.}
The observations of terrestrial thunderstorms indicate electric field strengths on the order of
$E_E \sim 10^5\, \mbox{V/m}$ and the lightning length scales are $L_E\sim 10^3 ~\mbox{m}$ \cite{NAP898}. 
Using the equation \eqref{necessary} for the mCP charge and mass,
we obtain for the Bose-enhancement case (cf.~\cite{Berlin:2024dwg})
\begin{equation}\label{constr_E_bose}
    \text{Bose enh.}~   ~~q_E \gtrsim 10^{-18},  ~~ m_E\lesssim10^{-10}~\mbox{eV}.
\end{equation} 

The constraints on fermionic mCP \eqref{master-no-bose} require lightning discharge characteristics. 
For typical terrestrial lightning, the peak current is $J_{E} = 3\times10^4$ A and $t_{E_l}\ \sim 10^{-2} \mbox{s}$ \footnote{Further in this section, the index ``l'' after the letter of the planet corresponds to ``lightning'', and the index ``p'' corresponds to ``pause''} \cite{heidler2008parameters, uman2007review}, and $t_{E_P} \simeq 1$ s \cite{WILLIAMS1999245}. The bound on the mCP charge \eqref{master-no-bose} reads, 
\begin{equation}\label{eq_noBose_E}
    \text{no Bose enh.}~   q_E \gtrsim 9.6 \times 10^{-7}, ~~ m_E \lesssim 1.4\times10^{-4}~\mbox{eV}.
\end{equation}

Now we consider how the Pauli principle affects fermions in the presence of a potential well as shown in Fig. \ref{fig:Clouds2}. First, we estimate the number of fermionic mCPs \eqref{eq:Fermi_number}  and the number necessary to discharge \eqref{eq:discharge number}. Taking the reference values \eqref{eq_noBose_E}, we obtain that 
the number of particles $N_F \simeq 3 \times 10^{13} $ that accumulate in the potential well is significantly less than the discharge number of particles $N_Q \simeq 6 \times 10^{24}$ in the cloud capacitor. Thus, the effect of Pauli blocking is in fact crucial for fermions in the presence of a cloud configuration (Fig. \ref{fig:Clouds2}). 
Solving $N_F=N_Q$ with respect to $q$, we obtain the constraint.
\begin{equation}
    q_E\gtrsim 2\times10^{-4},
\end{equation}
which is significantly worse than for non-layered structure \eqref{eq_noBose_E}. Further in the letter we provide the constraint of that type for Saturn. 

\textbf{Venus.}
 Lightnings on Venus have been detected by the planetary missions Venera-9 and Venera-10. These missions reported on the presence of a lightning area $S_V = 5 \times 10^{10}\; \mbox{m}^2$ in the atmosphere of Venus; the average length of lightning is $L_V \sim 10^4$ m, the duration of lightning is $t_{V_l} \sim 0.25$ s, and the number of lightning strikes per second is $\nu_V \sim 100\,s^{-1}$ \cite{krasnopolsky1980lightnings}. So, the pause time in a volume $L_V^3$ is $t_{V_p}=5$ s. 

The value of the electric field critical for the Venus atmosphere can be determined in two ways. First, three-dimensional modeling of the Venus atmosphere shows the critical value $74$ Td \cite{perez2017three}, which corresponds, using the ideal gas approximation, to $E_V \sim 3 \cdot 10^6$ V/m.

The laboratory experiment simulating the Venus atmosphere shows the critical field $E_V=1.9 \times 10^6\; \mbox{V/m}$ \cite{robledo2011space}. Both values of the electric field are of the same order of magnitude, but we consider \cite{robledo2011space} to be more precise. 

There are no direct measurements of the current of Venusian lightning; therefore, we will estimate the current using the charge density. The assumed charge density for lightning on Venus is estimated to be $\rho_V\sim 28\times10^{-9}~\mbox{C}/\mbox{m}^3$; the radius of the charge area is $R_{V_c}\sim 10^4~\mbox{m}$ \cite{KUMAR2021114473}. From the data from the measurements of the Venera 9 and Venera 10 satellites, we can obtain the current $J_V\simeq2\times10^5\mbox{A}$.


Taking into account these numerical values, we estimate constraints on the charge $q$ with and without Bose enhancement,
\begin{align}
  &\text{no Bose enh.}~ q_V \gtrsim 4 \times 10^{-8},~~ m_V\lesssim1.3\times10^{-4}~\mbox{eV}\\
  &\text{Bose enh.}~~   q_V \gtrsim 7 \times 10^{-22}, ~~ m_V \lesssim 2 \times10^{-11}\;\mbox{eV} .
\end{align}


\textbf{Saturn.}
The existence of lightning on Saturn was first observed by the Voyagers (SED), which were identified with lightnings 
after being further investigated by the Cassini mission \cite{dyudina2010detection, yair2008updated, fischer2008atmospheric}. 
The thunderstorm parameters are as follows: $E_S = 1.5 \times 10^7 ~\mbox{V/m}$ and $L_S \sim 10^5~\mbox{m}$ \cite{perez2017three}. 

As a result, we obtain the constraints for the charge  mass in the case of Bose-enhancement
\begin{equation}
 \text{Bose enh.}~   q_S \gtrsim 1.3 \times 10^{-24}, ~~ m_S \lesssim 2 \times 10^{-12}~\mbox{eV}.
\end{equation}

The obtained constraints are 6 orders of magnitude better than the Earth's \eqref{constr_E_bose}, since for Saturn both the length between clouds $L$ and electric field $E$ are more than one order of magnitude compared to Earth, see \eqref{necessary}.

In order to estimate the mCP constraint without Bose-enhancement, we use the charge momentum for the Saturn lightning $M_{Q_S} = 10^6\, \mbox{C}\cdot\mbox{km}$ \cite{DUBROVIN2014313, luque2014coupling}, which gives the charge $Q_S = M_{Q_S} /L_S = 10^4$ C.  The duration of lightning is $3\times10^{-2}$ s \cite{fischer2008atmospheric}, so the electric  current reads $J_S \sim 2.5 \times 10^5$ A. 
We estimate $t_{pause}$ using the radius of the lightning region $R_{S_l}=10^{6}~\mbox{m}$ \cite{fischer2008atmospheric} and the radius of the charged area $R_{S_c}=10^4~\mbox{m}$. The data about the frequency of lightning are described as several tens per second; therefore, we take an approximate value $\nu_{S}=50~\mbox{s}^{-1}$ \cite{fischer2008atmospheric}. So, the time of pause between lightning strikes is $t_{S_{p}}\sim200~\mbox{s}$.

The resulting mCP charge constraint is 
\begin{equation}
     \text{no Bose enh.}~   q_S \gtrsim 1.7 \times 10^{-10},  ~ ~m_S\lesssim2\times10^{-5}~\mbox{eV}.
\end{equation}
At last, we derive the constraints on fermionic mCPs assuming layered cloud structure and hence taking the account the Pauli blocking effect. For the corresponding calculations (see \eqref{eq:Fermi_number},\eqref{eq:discharge number}) we take the dielectric permittivity of Saturn’s atmosphere $\varepsilon\approx 2$  (see the refractive index in \cite{SANTER1981496, SANZREQUENA2018284}), obtaining 
\begin{equation}
    q_S \gtrsim 9 \times 10^{-5}.
\end{equation}
This value is still significantly less than other constraints so we further do not calculate it for other planets.

\textbf{Jupiter.}
The Jovian atmosphere has been explored by a variety of spacecraft, such as the Juno mission \cite{bolton2017juno}, Galileo measurements \cite{seiff1996structure}, and many others \cite{scarf1981upper, aglyamov2021lightning}. 

For Jovian lightning, several parameters have been measured: the typical lightning channel length is $L_J \sim 10^5\; \mbox{m}$ \cite{perez2017three} and the reduced electric field is $E_J/N_J\sim 100~\mbox{Td}$\footnote{$1~\mbox{Td}=10^{-21}\mbox{V/m}^2$.} \cite{luque2014coupling}. To convert the latter into an absolute electric field, we require atmospheric parameters at the altitude of approximately 85 km, where the pressure $P_J=1~\mbox{bar}$ and temperature $T_J \sim 165~\mbox{K}$ \cite{bagenal2004jupiter, nagy2008comparative}. Consequently, this yields a concentration $N_J \sim 4.4 \times 10^{25}~\mbox{m}^{-3}$ and electric field $E_J\sim4.4\times10^6~\mbox{V/m}$, consistent with model predictions \cite{YAIR1995421}.

Applying these parameters for mCP, we obtain charge and mass
\begin{equation}
\text{Bose enh.}~   q_J\gtrsim4.5\times10^{-24} , ~~ m_J\lesssim2\times10^{-12}~\mbox{eV}.
\end{equation}

\textbf{Titan (moon of Saturn).}
The Voyager 1 flyby of Titan in 1980 first raised the possibility of lightning on Saturn's largest moon. Later, during its flybys in 2005, Cassini observed several types of clouds in which lightning could potentially be produced \cite{porco2005imaging, griffith2006evidence}. 
In the models of Titan's atmosphere  \cite{TOKANO2001539, petculescu2014predicting} the maximum electric field is $E_T\sim2\times10^6~\mbox{V/m}$ and the possibility of lightning with a length of about $L_T\sim2\times10^4~\mbox{m}$. These parameters correspond to the constraints on the charge and mass of mCP
\begin{equation}
\text{Bose enh.}~   q_T\gtrsim2.5\times10^{-22} , ~~m_T\lesssim10^{-11}\mbox{eV}.
\end{equation}

\textbf{Ice Giants (Uranus and Neptune).}
In the Ice Giants' atmosphere, the radio-frequency signals identified with whistlers of lightning were  detected by Voyager 2 \cite{aplin2020atmospheric, aplin2017lightning}. In some theoretical models \cite{aplin2020atmospheric}, lightning strikes can occur in the cloud layers $H_2S-NH_3$ of Neptune \cite{GIBBARD1999227}. 

It is assumed that the length of lightning on Ice Giants will be longer than on Earth, but of the same order or slightly shorter than on Gas Giants, so we will roughly use the value of $L_{IG} \sim 10^5~\mbox{m}$. The required breakdown electric field for the $H_2N$ cloud corresponds to a pressure of 4 bar and a value of $E_{IG}\sim2\times10^6~\mbox{V/m}$.

The estimates for the constraints on the charge and mass of mCP are
\begin{equation}
\text{Bose enh.}~   q_{IG}\gtrsim10^{-23} , ~~m_{IG}\lesssim2\times10^{-12}~\mbox{eV}.
\end{equation}

The constraints for all considered planets are shown in figures \ref{fig_TS} and \ref{fig_TS_noBose}.

\section{Conclusion}\label{sec_conclusion}

In this letter, we set the constraints on the parameters of millicharged particles, both scalar or fermion, by the absence of supercritical Schwinger pair production of mCPs in thunderstorms in the atmospheres of Solar system planets.  Although the thunderstorms on Earth are closer to us and have been studied much more, the thunderstorms on gas giants are significantly more intense and extensive, so the corresponding constraints have improved by several orders of magnitude. 

Assuming the layered cloud structure, we obtain the  constraint for bosons that came from the thunderstorm in Saturn's atmosphere,
\begin{align}
\label{constraints}
    &q_S > 10^{-24}, \qquad \text{Bose-enhancement}.
\end{align}
These constraints on bosons are the best in the literature. Considering fermionic mCPs in the layered cloud structure, we must take into account the Pauli blocking effect, which made the obtained constraint significantly worser, 
\[
q_S>10^{-4}, \qquad \text{Pauli blocking}.
\]

If the clouds do not form the potential well, we obtain constraints on the charge from the Saturn's atmosphere
\[q_S>10^{-11}.\]

The constraints on bosonic mCPs are more than $10$ orders of magnitude better than those on fermion mCPs. The reason is the Bose enhancement. It can be realized only in the case of a ladder structure of clouds that form an electrostatic well for mCPs which seems to really exist in Saturn atmosphere \cite{fischer2008atmospheric}. We hope that a proper analysis of thunderstorms on Jupiter can further improve the constraints \eqref{constraints}. The constraints from Titan, ice giants, and Venus are worse than those for gas giants, as expected.
Moreover, more detailed study of planets in the Solar system may reveal cloud configurations where lightning strike should occur but in fact is absent. This could indicate that the clouds are discharged by another mechanism, such as mCPs.

The phenomenon investigated in this letter can be considered a providing example of how to use the physics of Solar system planets to probe some ``new physics'' hypotheses. For example, an investigation of the influence of axion-like-particles and dark photons may be an interesting task.

\paragraph{Funding} The work was supported by the grant of Russian Science Foundation № 25-22-00932.

\paragraph{Acknowledgements} Ekaterina Dmitrieva is a scholarship holder of the ``BASIS'' Foundation.

\bibliography{biblio}

@article{TOKANO2001539,
title = {Modelling of thunderclouds and lightning generation on Titan},
journal = {Planetary and Space Science},
volume = {49},
number = {6},
pages = {539-560},
year = {2001},
issn = {0032-0633},
doi = {https://doi.org/10.1016/S0032-0633(00)00170-7},
url = {https://www.sciencedirect.com/science/article/pii/S0032063300001707},
author = {T Tokano and G.J Molina-Cuberos and H Lammer and W Stumptner}
}

@article{perez2017three,
  title={Three-dimensional modeling of lightning-induced electromagnetic pulses on Venus, Jupiter, and Saturn},
  author={P{\'e}rez-Invern{\'o}n, Francisco J and Luque, Alejandro and Gordillo-V{\'a}zquez, Francisco J},
  journal={Journal of Geophysical Research: Space Physics},
  volume={122},
  number={7},
  pages={7636--7653},
  year={2017},
  publisher={Wiley Online Library}
}

@article{gierasch2000observation,
  title={Observation of moist convection in Jupiter's atmosphere},
  author={Gierasch, PJ and Ingersoll, AP and Banfield, D and Ewald, SP and Helfenstein, P and Simon-Miller, A and Vasavada, A and Breneman, HH and Senske, DA and Galileo Imaging Team and others},
  journal={Nature},
  volume={403},
  number={6770},
  pages={628--630},
  year={2000},
  publisher={Nature Publishing Group UK London}
}

@article{williams1983planetary,
  title={Planetary lightning: Earth, Jupiter, and Venus},
  author={Williams, Mark A and Krider, E Philip and Hunten, Donald M},
  journal={Reviews of Geophysics},
  volume={21},
  number={4},
  pages={892--902},
  year={1983},
  publisher={Wiley Online Library}
}

@article{Berlin:2024dwg,
    author = "Berlin, Asher and Harnik, Roni and Li, Ying-Ying and Xu, Bin",
    title = "{Millicharged Condensates on Earth}",
    eprint = "2404.16094",
    archivePrefix = "arXiv",
    primaryClass = "hep-ph",
    reportNumber = "FERMILAB-PUB-24-0002-SQMS-T, USTC-ICTS/PCFT-24-08",
    month = "4",
    year = "2024",
    journal = "",
    volume = ""
}

@BOOK{NAP898,
  author    = "National Research Council",
  title     = "The Earth;s Electrical Environment",
  isbn      = "978-0-309-03680-1",
  doi       = "10.17226/898",
  url       = "https://nap.nationalacademies.org/catalog/898/the-earths-electrical-environment",
  year      = 1986,
  publisher = "The National Academies Press",
  address   = "Washington, DC"
}

@article{luque2014coupling,
  title={Coupling between atmospheric layers in gaseous giant planets due to lightning-generated electromagnetic pulses},
  author={Luque, Alejandro and Dubrovin, D and Gordillo-V{\'a}zquez, Francisco J and Ebert, Ute and Parra-Rojas, Francisco C and Yair, Yoav and Price, C},
  journal={Journal of Geophysical Research: Space Physics},
  volume={119},
  number={10},
  pages={8705--8720},
  year={2014},
  publisher={Wiley Online Library}
}

@book{bagenal2004jupiter,
  title={Jupiter: The Planet, Satellites and Magnetosphere},
  author={Bagenal, F. and Dowling, T.E. and McKinnon, W.B.},
  number={2},
  isbn={9780521818087},
  lccn={2004040789},
  series={Cambridge Planetary Science},
  url={https://books.google.ru/books?id=8GcGRXlmxWsC},
  year={2004},
  publisher={Cambridge University Press}
}

@book{nagy2008comparative,
  title={Comparative Aeronomy},
  author={Nagy, A.F. and Balogh, A. and Cravens, T.E. and Mendillo, M. and M{\"u}ller-Wodarg, I.},
  isbn={9780387878256},
  lccn={2008939139},
  series={Space Sciences Series of ISSI},
  url={https://books.google.ru/books?id=ZT4pOCjjFq4C},
  year={2008},
  publisher={Springer New York}
}

@article{aplin2020atmospheric,
  title={Atmospheric electricity at the ice giants},
  author={Aplin, KL and Fischer, Georg and Nordheim, TA and Konovalenko, Alexandr and Zakharenko, Vyacheslav and Zarka, P},
  journal={Space Science Reviews},
  volume={216},
  number={2},
  pages={26},
  year={2020},
  publisher={Springer}
}

@article{bolton2017juno,
  title={The juno mission},
  author={Bolton, Scott J and Lunine, J and Stevenson, D and Connerney, JEP and Levin, S and Owen, TC and Bagenal, F and Gautier, D and Ingersoll, AP and Orton, GS and others},
  journal={Space Science Reviews},
  volume={213},
  number={1},
  pages={5--37},
  year={2017},
  publisher={Springer}
}

@article{aglyamov2021lightning,
  title={Lightning generation in moist convective clouds and constraints on the water abundance in Jupiter},
  author={Aglyamov, Yury S and Lunine, Jonathan and Becker, Heidi N and Guillot, Tristan and Gibbard, Seran G and Atreya, Sushil and Bolton, Scott J and Levin, Steven and Brown, Shannon T and Wong, Michael H},
  journal={Journal of Geophysical Research: Planets},
  volume={126},
  number={2},
  pages={e2020JE006504},
  year={2021},
  publisher={Wiley Online Library}
}

@article{dyudina2010detection,
  title={Detection of visible lightning on Saturn},
  author={Dyudina, Ulyana A and Ingersoll, AP and Ewald, SP and Porco, CC and Fischer, G and Kurth, WS and West, RA},
  journal={Geophysical Research Letters},
  volume={37},
  number={9},
  year={2010},
  publisher={Wiley Online Library}
}

@article{yair2008updated,
  title={Updated review of planetary atmospheric electricity},
  author={Yair, Yoav and Fischer, Georg and Sim{\~o}es, Fernando and Renno, N and Zarka, Philippe},
  journal={Space science reviews},
  volume={137},
  number={1},
  pages={29--49},
  year={2008},
  publisher={Springer}
}

@article{aplin2017lightning,
  title={Lightning detection in planetary atmospheres},
  author={Aplin, Karen L and Fischer, Georg},
  journal={Weather},
  volume={72},
  number={2},
  pages={46--50},
  year={2017},
  publisher={Wiley Online Library}
}

@article{GIBBARD1999227,
title = {Lightning on Neptune},
journal = {Icarus},
volume = {139},
number = {2},
pages = {227-234},
year = {1999},
issn = {0019-1035},
doi = {https://doi.org/10.1006/icar.1999.6101},
url = {https://www.sciencedirect.com/science/article/pii/S0019103599961018},
author = {S.G. Gibbard and E.H. Levy and J.I. Lunine and I. {de Pater}}
}

@article{HOLDOM1986196,
title = {Two U(1)'s and e charge shifts},
journal = {Physics Letters B},
volume = {166},
number = {2},
pages = {196-198},
year = {1986},
issn = {0370-2693},
doi = {https://doi.org/10.1016/0370-2693(86)91377-8},
url = {https://www.sciencedirect.com/science/article/pii/0370269386913778},
author = {Bob Holdom}
}

@article{Arza:2025cou,
    author = "Arza, Ariel and Gong, Yuanlin and Shu, Jing and Wu, Lei and Yuan, Qiang and Zhu, Bin",
    title = "{Geomagnetic Signal of Millicharged Dark Matter}",
    eprint = "2501.14949",
    archivePrefix = "arXiv",
    primaryClass = "hep-ph",
    month = "1",
    year = "2025",
    journal= ""
}

@mastersthesis{Lepidi:2007vnd,
    author = "Lepidi, Angela",
    title = "{Phenomenology and cosmology of millicharged particles and experimental prospects for their search}",
    eprint = "0809.4854",
    archivePrefix = "arXiv",
    primaryClass = "hep-ph",
    type = "Laurea thesis",
    school = "L'Aquila U.",
    year = "2007"
}

@article{IGNATIEV1979315,
title = {Is the electric charge conserved?},
journal = {Physics Letters B},
volume = {84},
number = {3},
pages = {315-318},
year = {1979},
issn = {0370-2693},
doi = {https://doi.org/10.1016/0370-2693(79)90048-0},
url = {https://www.sciencedirect.com/science/article/pii/0370269379900480},
author = {A.Yu. Ignatiev and V.A. Kuzmin and M.E. Shaposhnikov},
}

@article{Berlin:2020pey,
    author = "Berlin, Asher and Hook, Anson",
    title = "{Searching for Millicharged Particles with Superconducting Radio-Frequency Cavities}",
    eprint = "2001.02679",
    archivePrefix = "arXiv",
    primaryClass = "hep-ph",
    doi = "10.1103/PhysRevD.102.035010",
    journal = "Phys. Rev. D",
    volume = "102",
    number = "3",
    pages = "035010",
    year = "2020"
}

@article{Jaeckel:2021xyo,
    author = "Jaeckel, Joerg and Schenk, Sebastian",
    title = "{Challenging the Stability of Light Millicharged Dark Matter}",
    eprint = "2102.08394",
    archivePrefix = "arXiv",
    primaryClass = "hep-ph",
    reportNumber = "IPPP/20/81",
    doi = "10.1103/PhysRevD.103.103523",
    journal = "Phys. Rev. D",
    volume = "103",
    number = "10",
    pages = "103523",
    year = "2021"
}

@article{Okun:1982xi,
    author = "Okun, L. B.",
    title = "{LIMITS OF ELECTRODYNAMICS: PARAPHOTONS?}",
    reportNumber = "ITEP-48-1982",
    journal = "Sov. Phys. JETP",
    volume = "56",
    pages = "502",
    year = "1982"
}

@article{Batell:2005wa,
    author = "Batell, Brian and Gherghetta, Tony",
    title = "{Localized U(1) gauge fields, millicharged particles, and holography}",
    eprint = "hep-ph/0512356",
    archivePrefix = "arXiv",
    reportNumber = "UMN-TH-2425-05",
    doi = "10.1103/PhysRevD.73.045016",
    journal = "Phys. Rev. D",
    volume = "73",
    pages = "045016",
    year = "2006"
}

@article{Abel:2003ue,
    author = "Abel, S. A. and Schofield, B. W.",
    title = "{Brane anti-brane kinetic mixing, millicharged particles and SUSY breaking}",
    eprint = "hep-th/0311051",
    archivePrefix = "arXiv",
    reportNumber = "IPPP-03-69, DCPT-03-138",
    doi = "10.1016/j.nuclphysb.2004.02.037",
    journal = "Nucl. Phys. B",
    volume = "685",
    pages = "150--170",
    year = "2004"
}

@article{Abel:2008ai,
    author = "Abel, S. A. and Goodsell, M. D. and Jaeckel, J. and Khoze, V. V. and Ringwald, A.",
    title = "{Kinetic Mixing of the Photon with Hidden U(1)s in String Phenomenology}",
    eprint = "0803.1449",
    archivePrefix = "arXiv",
    primaryClass = "hep-ph",
    reportNumber = "IPPP-08-14, DESY-08-026",
    doi = "10.1088/1126-6708/2008/07/124",
    journal = "JHEP",
    volume = "07",
    pages = "124",
    year = "2008"
}

@article{Pati:1973uk,
    author = "Pati, Jogesh C. and Salam, Abdus",
    title = "{Unified Lepton-Hadron Symmetry and a Gauge Theory of the Basic Interactions}",
    reportNumber = "IC-73-41-INT-REP",
    doi = "10.1103/PhysRevD.8.1240",
    journal = "Phys. Rev. D",
    volume = "8",
    pages = "1240--1251",
    year = "1973"
}

@article{Georgi:1974my,
    author = "Georgi, Howard",
    editor = "Carlson, Hugh C. Carl E. Wolfe",
    title = "{The State of the Art{\textemdash}Gauge Theories}",
    doi = "10.1063/1.2947450",
    journal = "AIP Conf. Proc.",
    volume = "23",
    pages = "575--582",
    year = "1975"
}

@article{Preskill:1984gd,
    author = "Preskill, John",
    title = "{MAGNETIC MONOPOLES}",
    reportNumber = "CALT-68-1108",
    doi = "10.1146/annurev.ns.34.120184.002333",
    journal = "Ann. Rev. Nucl. Part. Sci.",
    volume = "34",
    pages = "461--530",
    year = "1984"
}

@article{Wen:1985qj,
    author = "Wen, Xiao-Gang and Witten, Edward",
    title = "{Electric and Magnetic Charges in Superstring Models}",
    reportNumber = "Print-85-0468 (PRINCETON)",
    doi = "10.1016/0550-3213(85)90592-9",
    journal = "Nucl. Phys. B",
    volume = "261",
    pages = "651--677",
    year = "1985"
}

@article{Shiu:2013wxa,
    author = "Shiu, Gary and Soler, Pablo and Ye, Fang",
    title = "{Milli-Charged Dark Matter in Quantum Gravity and String Theory}",
    eprint = "1302.5471",
    archivePrefix = "arXiv",
    primaryClass = "hep-th",
    reportNumber = "MAD-TH-13-01",
    doi = "10.1103/PhysRevLett.110.241304",
    journal = "Phys. Rev. Lett.",
    volume = "110",
    number = "24",
    pages = "241304",
    year = "2013"
}

@article{Brahm:1989jh,
    author = "Brahm, David E. and Hall, Lawrence J.",
    title = "{U(1)-prime DARK MATTER}",
    reportNumber = "LBL-27847, UCB-PTH-89/21",
    doi = "10.1103/PhysRevD.41.1067",
    journal = "Phys. Rev. D",
    volume = "41",
    pages = "1067",
    year = "1990"
}

@article{Feng:2009mn,
    author = "Feng, Jonathan L. and Kaplinghat, Manoj and Tu, Huitzu and Yu, Hai-Bo",
    title = "{Hidden Charged Dark Matter}",
    eprint = "0905.3039",
    archivePrefix = "arXiv",
    primaryClass = "hep-ph",
    reportNumber = "UCI-TR-2009-06",
    doi = "10.1088/1475-7516/2009/07/004",
    journal = "JCAP",
    volume = "07",
    pages = "004",
    year = "2009"
}

@article{Cline:2012is,
    author = "Cline, James M. and Liu, Zuowei and Xue, Wei",
    title = "{Millicharged Atomic Dark Matter}",
    eprint = "1201.4858",
    archivePrefix = "arXiv",
    primaryClass = "hep-ph",
    doi = "10.1103/PhysRevD.85.101302",
    journal = "Phys. Rev. D",
    volume = "85",
    pages = "101302",
    year = "2012"
}

@article{Davidson:2000hf,
    author = "Davidson, Sacha and Hannestad, Steen and Raffelt, Georg",
    title = "{Updated bounds on millicharged particles}",
    eprint = "hep-ph/0001179",
    archivePrefix = "arXiv",
    reportNumber = "CERN-TH-99-384",
    doi = "10.1088/1126-6708/2000/05/003",
    journal = "JHEP",
    volume = "05",
    pages = "003",
    year = "2000"
}

@article{Cohen:2008wz,
    author = "Cohen, Thomas D. and McGady, David A.",
    title = "{The Schwinger mechanism revisited}",
    eprint = "0807.1117",
    archivePrefix = "arXiv",
    primaryClass = "hep-ph",
    doi = "10.1103/PhysRevD.78.036008",
    journal = "Phys. Rev. D",
    volume = "78",
    pages = "036008",
    year = "2008"
}

@article{
Pioneer10,
author = {Albert G. Opp },
title = {Pioneer 10 Mission: Summary of Scientific Results from the Encounter with Jupiter},
journal = {Science},
volume = {183},
number = {4122},
pages = {302-303},
year = {1974},
doi = {10.1126/science.183.4122.302},
URL = {https://www.science.org/doi/abs/10.1126/science.183.4122.302},
eprint = {https://www.science.org/doi/pdf/10.1126/science.183.4122.302}}

@article{Berlin:2021kcm,
    author = "Berlin, Asher and Schutz, Katelin",
    title = "{Helioscope for gravitationally bound millicharged particles}",
    eprint = "2111.01796",
    archivePrefix = "arXiv",
    primaryClass = "hep-ph",
    reportNumber = "MIT-CTP/5358, FERMILAB-PUB-21-626-T",
    doi = "10.1103/PhysRevD.105.095012",
    journal = "Phys. Rev. D",
    volume = "105",
    number = "9",
    pages = "095012",
    year = "2022"
}

@article{Chang:2018rso,
    author = "Chang, Jae Hyeok and Essig, Rouven and McDermott, Samuel D.",
    title = "{Supernova 1987A Constraints on Sub-GeV Dark Sectors, Millicharged Particles, the QCD Axion, and an Axion-like Particle}",
    eprint = "1803.00993",
    archivePrefix = "arXiv",
    primaryClass = "hep-ph",
    reportNumber = "YITP-SB-18-01, FERMILAB-PUB-17-432-T",
    doi = "10.1007/JHEP09(2018)051",
    journal = "JHEP",
    volume = "09",
    pages = "051",
    year = "2018"
}

@article{Cruz:2022otv,
    author = "Cruz, Akaxia and McQuinn, Matthew",
    title = "{Astrophysical plasma instabilities induced by long-range interacting dark matter}",
    eprint = "2202.12464",
    archivePrefix = "arXiv",
    primaryClass = "astro-ph.CO",
    doi = "10.1088/1475-7516/2023/04/028",
    journal = "JCAP",
    volume = "04",
    pages = "028",
    year = "2023"
}

@article{schwinger1951gauge,
  title={On gauge invariance and vacuum polarization},
  author={Schwinger, Julian},
  journal={Physical Review},
  volume={82},
  number={5},
  pages={664},
  year={1951},
  publisher={APS}
}

@article{Dmitrieva:2025ohn,
    author = "Dmitrieva, Ekaterina and Satunin, Petr",
    title = "{Resonant production of millicharged scalars in k{\textasciicircum}2 {\ensuremath{>}} 0 electromagnetic wave background}",
    eprint = "2510.25505",
    archivePrefix = "arXiv",
    primaryClass = "hep-ph",
    reportNumber = "INR-TH-2025-019",
    month = "10",
    year = "2025",
    journal = " "
}

@article{doi:10.1142/S0217732314500540,
author = {Li, Xin and Voloshin, M. B.},
title = {Electric discharge in vacuum by minicharged particles},
journal = {Modern Physics Letters A},
volume = {29},
number = {11},
pages = {1450054},
year = {2014},
doi = {10.1142/S0217732314500540},
URL = {https://doi.org/10.1142/S0217732314500540},
eprint = {https://doi.org/10.1142/S0217732314500540}
}

@article{klein1978bose,
  title={Bose condensation in supercritical external fields charged condensates},
  author={Klein, Abraham and Rafelski, Johann},
  journal={Zeitschrift f{\"u}r Physik A Atoms and Nuclei},
  volume={284},
  number={1},
  pages={71--81},
  year={1978},
  publisher={Springer}
}

@article{NIKISHOV1970346,
title = {Barrier scattering in field theory removal of Klein paradox},
journal = {Nuclear Physics B},
volume = {21},
number = {2},
pages = {346-358},
year = {1970},
issn = {0550-3213},
doi = {https://doi.org/10.1016/0550-3213(70)90527-4},
url = {https://www.sciencedirect.com/science/article/pii/0550321370905274},
author = {A.I Nikishov}
}

@article{Kouvaris:2025tom,
    author = "Kouvaris, Chris and Shoemaker, Ian M.",
    title = "{Millicharged Particle Production in Pulsars via the Schwinger Effect}",
    eprint = "2511.04763",
    archivePrefix = "arXiv",
    primaryClass = "hep-ph",
    month = "11",
    year = "2025",
    journal = "",
    volume = ""
}

@article{Gninenko:2025aek,
    author = "Gninenko, Sergei N. and Krasnikov, N. V. and Kuleshov, Sergey and Lyubovitskij, Valery E. and Crivelli, P. and Kirpichnikov, D. V. and Bueno, L. Molina and Zhevlakov, Alexey S. and Sieber, H. and Voronchikhin, I. V.",
    title = "{Probing millicharged particles with NA64$\mu $ and LDMX}",
    eprint = "2505.04295",
    archivePrefix = "arXiv",
    primaryClass = "hep-ph",
    doi = "10.1140/epjc/s10052-026-15307-w",
    journal = "Eur. Phys. J. C",
    volume = "86",
    number = "2",
    pages = "140",
    year = "2026"
}

@article{wallace1974thermal,
  title={The thermal structure of the atmosphere of Jupiter},
  author={Wallace, L and Prather, Michael and Belton, MJ},
  journal={The Astrophysical Journal},
  volume={193},
  pages={481},
  year={1974}
}

@article{cook1979first,
  title={First results on Jovian lightning},
  author={Cook, AF and Duxbury, TC and Hunt, GE and others},
  journal={Nature},
  volume={280},
  year={1979}
}

@unpublished{JUNOwebsite,
    note = {https://www.missionjuno.swri.edu/}
}

@article{deMontigny:2023qft,
    author = "de Montigny, Marc and Ouimet, Pierre-Philippe A. and Pinfold, James and Shaa, Ameir and Staelens, Michael",
    title = "{Minicharged Particles at Accelerators: Progress and Prospects}",
    eprint = "2307.07855",
    archivePrefix = "arXiv",
    primaryClass = "hep-ph",
    month = "7",
    year = "2023",
    journal = ""
}

@article{Fiorillo:2024upk,
    author = "Fiorillo, Damiano F. G. and Vitagliano, Edoardo",
    title = "{Self-Interacting Dark Sectors in Supernovae Can Behave as a Relativistic Fluid}",
    eprint = "2404.07714",
    archivePrefix = "arXiv",
    primaryClass = "hep-ph",
    doi = "10.1103/PhysRevLett.133.251004",
    journal = "Phys. Rev. Lett.",
    volume = "133",
    number = "25",
    pages = "251004",
    year = "2024"
}

@article{becker2020small,
  title={Small lightning flashes from shallow electrical storms on Jupiter},
  author={Becker, Heidi N and Alexander, James W and Atreya, Sushil K and Bolton, Scott J and Brennan, Martin J and Brown, Shannon T and Guillaume, Alexandre and Guillot, Tristan and Ingersoll, Andrew P and Levin, Steven M and others},
  journal={Nature},
  volume={584},
  number={7819},
  pages={55--58},
  year={2020},
  publisher={Nature Publishing Group UK London}
}

@article{scarf1981upper,
  title={An upper bound to the lightning flash rate in Jupiter's atmosphere},
  author={Scarf, FL and Gurnett, DA and Kurth, WS and Anderson, RR and Shaw, RR},
  journal={Science},
  volume={213},
  number={4508},
  pages={684--685},
  year={1981},
  publisher={American Association for the Advancement of Science}
}

@article{kohlhase1977voyager,
  title={Voyager mission description},
  author={Kohlhase, CE and Penzo, Paul Anthony},
  journal={Space science reviews},
  volume={21},
  number={2},
  pages={77--101},
  year={1977},
  publisher={Springer}
}

@inproceedings{heidler2008parameters,
  title={Parameters of lightning current given in IEC 62305-background, experience and outlook},
  author={Heidler, Fridolin and Zischank, W and Flisowski, Z and Bouquegneau, Christian and Mazzetti, C},
  booktitle={29th International Conference on Lightning Protection},
  volume={23},
  pages={26},
  year={2008}
}

@article{uman2007review,
  title={A review of natural lightning: Experimental data and modeling},
  author={Uman, Martin A and Krider, E Philip},
  journal={IEEE Transactions on electromagnetic compatibility},
  number={2},
  pages={79--112},
  year={2007},
  publisher={IEEE}
}

@article{WILLIAMS1999245,
title = {The behavior of total lightning activity in severe Florida thunderstorms},
journal = {Atmospheric Research},
volume = {51},
number = {3},
pages = {245-265},
year = {1999},
issn = {0169-8095},
doi = {https://doi.org/10.1016/S0169-8095(99)00011-3},
url = {https://www.sciencedirect.com/science/article/pii/S0169809599000113},
author = {Earle Williams and Bob Boldi and Anne Matlin and Mark Weber and Steve Hodanish and Dave Sharp and Steve Goodman and Ravi Raghavan and Dennis Buechler},
}

@article{robledo2011space,
  title={Space charge effects and arc properties of simulated lightning on Venus},
  author={Robledo-Martinez, A and Sobral, H and Ruiz-Meza, A},
  journal={Journal of Geophysical Research: Space Physics},
  volume={116},
  number={A6},
  year={2011},
  publisher={Wiley Online Library}
}

@article{KUMAR2021114473,
title = {Venus lightning: Estimation of charge and dimensions of charge regions for lightning initiation},
journal = {Icarus},
volume = {365},
pages = {114473},
year = {2021},
issn = {0019-1035},
doi = {https://doi.org/10.1016/j.icarus.2021.114473},
url = {https://www.sciencedirect.com/science/article/pii/S0019103521001548},
author = {V.R. Dinesh Kumar and Jayesh P. Pabari and Kinsuk Acharyya and C.T. Russell},
}

@article{fischer2008atmospheric,
  title={Atmospheric electricity at Saturn},
  author={Fischer, Georg and Gurnett, Donald A and Kurth, William S and Akalin, Ferzan and Zarka, Philippe and Dyudina, Ulyana A and Farrell, William M and Kaiser, Michael L},
  journal={Space Science Reviews},
  volume={137},
  number={1},
  pages={271--285},
  year={2008},
  publisher={Springer}
}

@article{DUBROVIN2014313,
title = {Impact of lightning on the lower ionosphere of Saturn and possible generation of halos and sprites},
journal = {Icarus},
volume = {241},
pages = {313-328},
year = {2014},
issn = {0019-1035},
doi = {https://doi.org/10.1016/j.icarus.2014.06.025},
url = {https://www.sciencedirect.com/science/article/pii/S0019103514003455},
author = {D. Dubrovin and A. Luque and F.J. Gordillo-Vazquez and Y. Yair and F.C. Parra-Rojas and U. Ebert and C. Price},
}

@article{YAIR1995421,
title = {Lightning Generation in a Jovian Thundercloud: Results from an Axisymmetric Numerical Cloud Model},
journal = {Icarus},
volume = {115},
number = {2},
pages = {421-434},
year = {1995},
issn = {0019-1035},
doi = {https://doi.org/10.1006/icar.1995.1108},
url = {https://www.sciencedirect.com/science/article/pii/S0019103585711086},
author = {Yoav Yair and Zev Levin and Shalva Tzivion},
}

@article{Davidson:1993sj,
    author = "Davidson, Sacha and Peskin, Michael E.",
    title = "{Astrophysical bounds on millicharged particles in models with a paraphoton}",
    eprint = "hep-ph/9310288",
    archivePrefix = "arXiv",
    reportNumber = "SLAC-PUB-6360, CFPA-93-TH-31",
    doi = "10.1103/PhysRevD.49.2114",
    journal = "Phys. Rev. D",
    volume = "49",
    pages = "2114--2117",
    year = "1994"
}

@article{
doi:10.1126/science.212.4491.239,
author = {J. W. Warwick  and J. B. Pearce  and D. R. Evans  and T. D. Carr  and J. J. Schauble  and J. K. Alexander  and M. L. Kaiser  and M. D. Desch  and M. Pedersen  and A. Lecacheux  and G. Daigne  and A. Boischot  and C. H. Barrow },
title = {Planetary Radio Astronomy Observations from Voyager 1 Near Saturn},
journal = {Science},
volume = {212},
number = {4491},
pages = {239-243},
year = {1981},
doi = {10.1126/science.212.4491.239},
URL = {https://www.science.org/doi/abs/10.1126/science.212.4491.239},
eprint = {https://www.science.org/doi/pdf/10.1126/science.212.4491.239},
}

@article{ksanfomaliti1980discovery,
  title={Discovery of frequent lightning discharges in clouds on Venus},
  author={Ksanfomaliti, LV},
  journal={Nature},
  volume={284},
  number={5753},
  pages={244--246},
  year={1980},
  publisher={Nature Publishing Group UK London}
}

@techreport{ksanfomaliti1979electrical,
  title={Electrical discharges in the atmosphere of Venus},
  author={Ksanfomaliti, LV and Vasilchikov, NM and Ganpantserova, OF and Petrova, YV and Suvorov, AP and Filippov, GF and Yablonskaya, OV and Yabrova, LV},
  year={1979}
}

@article{krasnopolsky1980lightnings,
  title={On lightnings in the Venus atmosphere according to the Venera 9 and 10 data},
  author={Krasnopolsky, VA},
  journal={Cosmic Res},
  volume={18},
  number={556},
  pages={22},
  year={1980}
}

@article{HANSELL1995345,
title = {Optical Detection of Lightning on Venus},
journal = {Icarus},
volume = {117},
number = {2},
pages = {345-351},
year = {1995},
issn = {0019-1035},
doi = {https://doi.org/10.1006/icar.1995.1160},
url = {https://www.sciencedirect.com/science/article/pii/S0019103585711608},
author = {S.A. Hansell and W.K. Wells and D.M. Hunten},
}

@article{porco2005imaging,
  title={Imaging of Titan from the Cassini spacecraft},
  author={Porco, Carolyn C and Baker, Emily and Barbara, John and Beurle, Kevin and Brahic, Andre and Burns, Joseph A and Charnoz, Sebastien and Cooper, Nick and Dawson, Douglas D and Del Genio, Anthony D and others},
  journal={Nature},
  volume={434},
  number={7030},
  pages={159--168},
  year={2005},
  publisher={Nature Publishing Group UK London}
}

@article{griffith2006evidence,
  title={Evidence for a polar ethane cloud on Titan},
  author={Griffith, CA and Penteado, P and Rannou, Pascal and Brown, R and Boudon, Vincent and Baines, KH and Clark, R and Drossart, P and Buratti, B and Nicholson, P and others},
  journal={Science},
  volume={313},
  number={5793},
  pages={1620--1622},
  year={2006},
  publisher={American Association for the Advancement of Science}
}

@article{petculescu2014predicting,
  title={Predicting the characteristics of thunder on Titan: A framework to assess the detectability of lightning by acoustic sensing},
  author={Petculescu, Andi and Kruse, Roland},
  journal={Journal of Geophysical Research: Planets},
  volume={119},
  number={10},
  pages={2167--2176},
  year={2014},
  publisher={Wiley Online Library}
}

@article{Burrage:2009yz,
    author = "Burrage, C. and Jaeckel, J. and Redondo, J. and Ringwald, A.",
    title = "{Late time CMB anisotropies constrain mini-charged particles}",
    eprint = "0909.0649",
    archivePrefix = "arXiv",
    primaryClass = "astro-ph.CO",
    reportNumber = "DESY-09-132, DCPT-09-124, IPPP-09-62",
    doi = "10.1088/1475-7516/2009/11/002",
    journal = "JCAP",
    volume = "11",
    pages = "002",
    year = "2009"
}

@article{seiff1985models,
  title={Models of the structure of the atmosphere of Venus from the surface to 100 kilometers altitude},
  author={Seiff, Alvin and Schofield, JT and Kliore, AJ and Taylor, FW and Limaye, SS and Revercomb, HE and Sromovsky, LA and Kerzhanovich, VV and Moroz, VI and Marov, M Ya},
  journal={Advances in Space Research},
  volume={5},
  number={11},
  pages={3--58},
  year={1985},
  publisher={Elsevier}
}

@article{seiff1996structure,
  title={Structure of the atmosphere of Jupiter: Galileo probe measurements},
  author={Seiff, Alvin and Kirk, Donn B and Knight, Tony CD and Mihalov, John D and Blanchard, Robert C and Young, Richard E and Schubert, Gerald and Von Zahn, Ulf and Lehmacher, Gerald and Milos, Frank S and others},
  journal={Science},
  volume={272},
  number={5263},
  pages={844--845},
  year={1996},
  publisher={American Association for the Advancement of Science}
}

@article{aplin2017electrical,
  title={Electrical processes in planetary atmospheres},
  author={Aplin, Karen and Airey, Martin and Warriner-Bacon, Elliot},
  journal={arXiv preprint arXiv:1705.05597},
  year={2017}
}

@article{scarf1980lightning,
  title={Lightning on Venus: Orbiter detection of whistler signals},
  author={Scarf, FL and Taylor, WWL and Russell, CT and Brace, LH},
  journal={Journal of Geophysical Research: Space Physics},
  volume={85},
  number={A13},
  pages={8158--8166},
  year={1980},
  publisher={Wiley Online Library}
}

@article{wong2026radio,
  title={Radio pulse power distribution of lightning in Jupiter's 2021--2022 stealth superstorms},
  author={Wong, Michael H and Kolma{\v{s}}ov{\'a}, Ivana and Oyafuso, Fabiano A and Imai, Masafumi and Mizumoto, Shinji and Levin, Steven M and Sankar, Ramanakumar G and Simon, Amy A and Brueshaber, Shawn and Orton, Glenn S and others},
  journal={AGU Advances},
  volume={7},
  number={2},
  pages={e2025AV002083},
  year={2026},
  publisher={Wiley Online Library}
}

@book{landau2013statistical,
  title={Statistical physics: volume 5},
  author={Landau, Lev Davidovich and Lifshitz, Evgenii Mikhailovich},
  volume={5},
  year={2013},
  publisher={Elsevier}
}

@article{Gong:2025xsd,
    author = "Gong, Yuanlin and Tian, Hongliang and Wu, Lei and Zhu, Bin",
    title = "{Probing millicharged dark matter with magnetometer coupled to circuit}",
    eprint = "2504.09630",
    archivePrefix = "arXiv",
    primaryClass = "hep-ph",
    reportNumber = "CPTNP-2025-002",
    doi = "10.1007/s11433-025-2717-5",
    journal = "Sci. China Phys. Mech. Astron.",
    volume = "68",
    number = "8",
    pages = "280412",
    year = "2025"
}

@book{Rubakov:2017xzr,
    author = "Rubakov, Valery A. and Gorbunov, Dmitry S.",
    title = "{Introduction to the Theory of the Early Universe}: {Hot big bang theory}",
    doi = "10.1142/10447",
    isbn = "978-981-320-987-9, 978-981-320-988-6, 978-981-322-005-8",
    publisher = "World Scientific",
    address = "Singapore",
    year = "2017"
}

@article{Korwar:2017dio,
    author = "Korwar, Mrunal and Thalapillil, Arun M.",
    title = "{Novel Astrophysical Probes of Light Millicharged Fermions through Schwinger Pair Production}",
    eprint = "1709.07888",
    archivePrefix = "arXiv",
    primaryClass = "hep-ph",
    doi = "10.1007/JHEP04(2019)039",
    journal = "JHEP",
    volume = "04",
    pages = "039",
    year = "2019"
}

@article{SANTER1981496,
title = {Optical reflectance polarimetry of Saturn's globe and rings: IV. Aerosols in the upper atmosphere of Saturn},
journal = {Icarus},
volume = {48},
number = {3},
pages = {496-518},
year = {1981},
issn = {0019-1035},
doi = {https://doi.org/10.1016/0019-1035(81)90060-9},
url = {https://www.sciencedirect.com/science/article/pii/0019103581900609},
author = {Richard Santer and Audouin Dollfus},
}

@article{SANZREQUENA2018284,
title = {Haze and cloud structure of Saturn's North Pole and Hexagon Wave from Cassini/ISS imaging},
journal = {Icarus},
volume = {305},
pages = {284-300},
year = {2018},
issn = {0019-1035},
doi = {https://doi.org/10.1016/j.icarus.2017.12.043},
url = {https://www.sciencedirect.com/science/article/pii/S0019103517302920},
author = {J.F. Sanz-Requena and S. Pérez-Hoyos and A. Sánchez-Lavega and A. Antuñano and Patrick G.J. Irwin},
}

\end{document}